\documentclass[useAMS,usenatbib,letterpaper]{mn2e}

\usepackage{times}

\usepackage{graphicx}
\usepackage{amsmath}
\usepackage{amssymb}
\usepackage{natbib}
\usepackage[lowtilde]{url}

\newcommand{\apj}{Ap.J.}

\newcommand{\be}{\begin{equation}}
\newcommand{\ee}{\end{equation}}

\def\trh0{t_{rh}(0)}

\newcommand{\msun}{M_{\odot}}
\def\apgt{\ {\raise-.5ex\hbox{$\buildrel>\over\sim$}}\ }
\def\aplt{\ {\raise-.5ex\hbox{$\buildrel<\over\sim$}}\ }

\title[No Energy Equipartition in Globular Clusters] {No
  Energy Equipartition in Globular Clusters
  }\author[Trenti \& van der Marel] {Michele
  Trenti$^{1}$\thanks{E-mail addresses:
    trentiast.cam.ac.uk} and Roeland van der Marel$^{2}$\\
  $^{1}$Institute of Astronomy, University of Cambridge, United
  Kingdom \\ $^{2}$Space Telescope Science Institute, Baltimore, MD
  21210, USA}

\begin{document}

\date{.}

\pagerange{\pageref{firstpage}--\pageref{lastpage}} \pubyear{2013}

\maketitle

\label{firstpage}

\begin{abstract}
  It is widely believed that globular clusters evolve over many
  two-body relaxation times toward a state of energy equipartition, so
  that velocity dispersion scales with stellar mass as $\sigma \propto
  m^{-\eta}$ with $\eta = 0.5$. We show here that this is incorrect,
  using a suite of direct N-body simulations with a variety of
  realistic IMFs and initial conditions. No simulated system ever
  reaches a state close to equipartition. Near the center, the
  luminous main-sequence stars reach a maximum $\eta_{\rm max} \approx
  0.15 \pm 0.03$. At large times, all radial bins convergence on an
  asymptotic value $\eta_{\infty} \approx 0.08 \pm 0.02$. The
  development of this ``partial equipartition'' is strikingly similar
  across our simulations, despite the range of different initial
  conditions employed. Compact remnants tend to have higher $\eta$
  than main-sequence stars (but still $\eta < 0.5$), due to their
  steeper (evolved) mass function. The presence of an
  intermediate-mass black hole (IMBH) decreases $\eta$, consistent
  with our previous findings of a quenching of mass segregation under
  these conditions. All these results can be understood as a
  consequence of the Spitzer instability for two-component systems,
  extended by Vishniac to a continuous mass spectrum. Mass segregation
  (the tendency of heavier stars to sink toward the core) has often
  been studied observationally, but energy equipartition has not. Due
  to the advent of high-quality proper motion datasets from the Hubble
  Space Telescope, it is now possible to measure $\eta$ for real
  clusters.  Detailed data-model comparisons open up a new
  observational window on globular cluster dynamics and evolution. A
  first comparison of our simulations to observations of Omega Cen
  yields good agreement, supporting the view that globular clusters
  are not generally in energy equipartition. Modeling techniques that
  assume equipartition by construction (e.g., multi-mass Michie-King
  models) are approximate at best.
\end{abstract}




\section{Introduction}
\label{sec:intro}

Gravitational encounters within stellar systems in virial
equilibrium, such as globular clusters, drive evolution over the
two-body relaxation timescale. The evolution is toward a thermal
velocity distribution, in which stars of different mass have the same
energy \citep{spitzer87}. This thermalization also induces mass
segregation. As the system evolves toward energy equipartition, high
mass stars lose energy, decrease their velocity dispersion and tend to
sink toward the central regions. The opposite happens for low mass
stars, which gain kinetic energy, tend to migrate toward the outer
parts of the system, and preferentially escape the system in the
presence of a tidal field.

As first pointed out by \citet{spitzer69}, not all self-gravitating
systems can attain complete energy equipartition, i.e., a
mass-dependent velocity dispersion scaling as $\sigma(m)\propto
m^{-0.5}$. For example, in a simple two-component model with light and
heavy stars of masses $m_1$ and $m_2$, respectively, equipartition is
possible only if the mass fraction in heavy particles is smaller than
a critical value: $M_2/M_1 \lesssim 0.16 (m_1/m_2)^{3/2}$, where $M_1$
and $M_2$ are the total masses in stars of mass $m_1$ and $m_2$,
respectively \citep{spitzer69}. With too many massive particles, no
equipartition is possible, and the heavy particles tend to become a
self-gravitating system, producing a deep core collapse while
continuing to transfer energy to the lighter stars.

Subsequent research has refined our knowledge of the conditions
required to establish energy equipartition. This was studied for
two-component systems using a variety of numerical methods, including
pioneering work by \citet{spitzer_hart71} and \citet{inagaki84} with
Fokker-Planck and Monte Carlo schemes, followed by more realistic
simulations both with a Monte Carlo code \citep{watters00,fregeau02}
and with direct N-body integration
\citep{khalisi07}. \citet{vishniac78} generalized the analytical
Spitzer instability analysis to include a continuous mass spectrum. He
showed that clusters with realistic mass functions, such those of the
\citet{salp} form, are {\it unable to attain energy equipartition.}
With a complementary approach based on multi-component King
distribution function modeling, \citet{kondratev} showed that even if
a system attains equipartition in the energy phase space, the
mass-dependent kinematics does not necessarily scale as
$\sigma(m)\propto m^{-0.5}$.

Much of the theoretical interest in equipartition has focused on how
energy exchange affects the process and time scale of core collapse in
globular clusters. Both analytically and numerically, it has been
established that the core collapse timescale in a two-component model
is inversely proportional to the ratio of the heavy to light particle
masses (e.g., \citealt{portegies-zwart00,khalisi07}). This has the
important consequence that young star clusters can undergo a runaway
central collapse of massive stars within the first few million years
after their formation, possibly providing a pathway to the formation
of intermediate-mass black holes (IMBHs) \citep{portegies-zwart2004}.

One of the directly observable consequences of the drive towards
energy equipartition is spatial mass segregation. In simulations, the
amount of mass segregation reaches a steady-state configuration within
a few initial relaxation times \citep{gill08}. The presence of an
IMBH, or of a stellar-mass black hole binary, in the core of a stellar
cluster reduces the amount of mass segregation that develops
\citep{bau04,tre07b,gill08,umbreit12}.

Mass segregation has been measured in several globular clusters using
high-quality Hubble Space Telescope (HST) data. This only requires
observations of stellar positions and luminosities (on the main
sequence, luminosity correlates with mass). Such observations have
confirmed the qualitative picture that massive stars are
preferentially found in the core
\citep{demarchi1994,andreuzzi2000,paust2010}. However, interpretation
of such data requires detailed dynamical models. These are often
constructed {\it assuming energy equipartition} for all components
through Michie-King models
\citep[e.g.,][]{gunn,meylan1988,demarchi2000}. The accuracy of this is
unclear, given the theoretical results described above which indicate
that globular clusters may not generally be in energy equipartition.

A preferred approach is to compare data directly to the results of
numerical simulations. Detailed, data-model comparisons for globular
clusters such as NGC2298 and M10 have shown a remarkable quantitative
agreement with the mass segregation predictions from N-body
simulations \citep{pas09,beccari,umbreit12}\footnote{Note that these
  N-body models are to some degree approximate: They either have a
  smaller number of particles compared to stars for direct N-body
  integration \citep{pas09,beccari}, or resort to a Monte Carlo
  treatment of particle orbits \citep{umbreit12}.}. However, such
comparisons are challenging to carry out for general datasets. It not
only requires detailed numerical simulations, but the simulations must
also be ``observed'' to mimic the magnitude limits, completeness and
crowding effects in the data. These observational effects generally
vary with radial distance within a system, and between systems.

Mass segregation provides only an {\it indirect} measure of energy
equipartition: stellar energies can only be estimated statistically
from stellar positions, and this requires assumed knowledge about the
dynamical state and gravitational potential of the system. It has not
been possible historically to measure $\sigma(m)$ in globular clusters
{\it directly} from knowledge of individual stellar
velocities. Line-of-sight velocities of stars can be measured from
spectra, but those are difficult to obtain for faint stars below the
main-sequence turn-off. Stars in globular clusters with measured
line-of-sight velocities are therefore generally giant stars. These
stars all have very similar mass (because stellar evolution proceeds
rapidly after the main-sequence turn-off), making it impossible to
determine how the velocity dispersion $\sigma$ in a cluster depends on
stellar mass $m$.

This situation is now changing due to the advent of HST studies of
stellar {\it proper motions} in globular clusters, which can quantify
the kinematics of the stars as function of their position along the
main sequence (i.e., their mass). \citet{anderson} presented
measurements for more than 100,000 stars in the central region of
Omega Cen, finding $\sigma \propto m^{-0.2}$. Proper motion catalogs
for some two dozen additional clusters are under constructions
\citep{bellini}. This opens up a new observational window into the
dynamics of globular clusters.

To understand and exploit these new data there is a need for new
analysis of models to predict in detail the mass- and
spatially-dependent kinematics of globular clusters as a function of
initial conditions and evolutionary state. With new data and new
analysis in hand, there is the potential to validate our understanding
of globular cluster structure and evolution at a new level. Moreover,
knowledge of the $\sigma(m)$ relation may constrain structural
parameters that are difficult to assess otherwise, such as the
possible presence of a central IMBH.

The goal of this paper is to make the first step towards modeling
mass-dependent kinematics in the era of detailed proper motions
measurements. We analyze a sample of realistic direct N-body
simulations of star cluster evolution, and we also use the models to
interpret the specific observational results obtained for Omega Cen
(after rescaling model relaxation times to the one of Omega Cen to
account for the larger number of stars compared to the particles we
can simulate). The structure of the paper is as follows. In
Section~\ref{sec:sim} we introduce the numerical simulations that we
have carried out, and we describe how we analyze simulation snapshots
to construct $\sigma(m)$. The results of our analysis are presented in
Section~\ref{sec:results}. We compare the model predictions to Omega
Cen observations in Section~\ref{sec:omegacen}, and find excellent
consistency between models and observations. We summarize our key
findings and conclude with an outlook for future studies in
Section~\ref{sec:con}.

\begin{table*}
\begin{minipage}{110mm}
\caption{Summary of N-body simulations and energy equipartition results\label{tab:sim}}
\begin{tabular}{cccc|cccc}
  \hline
  \hline
  $N$ & $f$ & IMF & $W_0$ &  $t_{\rm half}/t_{rh}(0)$ & $t_{\rm max}/t_{rh}(0)$ &$\eta_{\rm max}$ & $\eta_{\infty}$ \\ 
(1) & (2) & (3) & (4) & (5) & (6) & (7) & (8) \\
  \hline
65536 & 0.00& MS & 3 &  1.0 &  7.2 &       $0.124\pm   0.019 $    & $0.072 \pm 0.022$ \\
65536 & 0.02& MS & 3 &  1.5&      4.7&      $ 0.124  \pm 0.024$ &    $  0.063 \pm    0.026$ \\
65536 & 0.055& MS & 3 &1.1 &     4.9 &     $ 0.119  \pm     0.018$&     $  0.075 \pm      0.016$\\
65536$^*$ & 0.00& MS & 5 &   0.9  &     3.4&     $ 0.144 \pm      0.006$&      $ 0.085 \pm      0.006$\\
65536 & 0.02& MS & 5 &  1.0   &    3.9 &      $0.141 \pm      0.008$ &   $    0.084 \pm      0.008$\\
65536 & 0.00& Sal & 7 &  1.3 &       6.2 &    $ 0.150 \pm      0.004$&   $    0.098 \pm      0.012$\\
32768 & 0.00& MS & 7 &  0.6 &       2.9 &      $  0.150  \pm 0.008$& $     0.094 \pm      0.006$ \\
32768 & 0.00& MS & 7 &  0.5 &       2.4 &      $  0.159  \pm     0.007$& $     0.078 \pm      0.011$ \\
32768 & 0.00& MS & 7 &  0.6 &       2.8 &       $0.158  \pm     0.008 $&    $  0.086  \pm     0.009$\\
32768 & 0.01& MS & 7 & 0.6 &       2.7&       $0.156 \pm      0.007 $   &  $ 0.086   \pm    0.011$\\
32768 & 0.03& MS & 7 & 0.5&     2.8 &      $ 0.134 \pm       0.008 $&    $  0.088 \pm      0.007$\\
32768 & 0.05& MS & 7 & 0.5&       2.3&      $ 0.137 \pm      0.009$&     $  0.076 \pm      0.016$\\
32768 & 0.10& MS & 7 &  0.5 &       2.1&       $0.127 \pm      0.008 $&     $ 0.069  \pm     0.007$\\
32768 & 0.00& Sa & 7 &   0.8  &     4.9&    $   0.151  \pm     0.007$&     $   0.090  \pm     0.011$\\
32768$^{\dag}$ & 0.00& MS & 7 &   0.5 &       2.2    &  $ 0.180    \pm   0.011   $&$    0.085 \pm      0.012$\\
32768$^{\dag}$ & 0.00& Sa & 7 &   0.5 &     2.6& $       0.181  \pm 0.010 $  &  $  0.101 \pm      0.009$ \\
32769$^{\ddag}$ & 0.00& MS & 7 & 0.8 &      4.4 &      $ 0.120 \pm      0.010 $    &$  0.097 \pm      0.007$\\
32768$^{\S}$ & 0.00& MS & 3 & 1.6 &    6.3&      $ 0.146 \pm      0.008 $&     $ 0.084  \pm     0.011$\\
\hline 
\end{tabular}
\medskip

{N-body simulations of star clusters in a tidal field with self
  consistent King model initial conditions. Column~(1): number of
  particles $N$. Column (2): binary fraction $f$. Column~(3): initial
  mass function used (Sa: Salpeter power law; MS: Miller \&
  Scalo). Column~(4): initial concentration of the density profile
  (King index $W_0$); Columns~(5)--(8): equipartition results for
  single main-sequence stars, quantified as described in
  Section~\ref{subsec:canondep}; Column~(5): time $t_{\rm half}$ to
  reach half of the maximum equipartition; Column~(6): time $t_{\rm
    max}$ to reach the maximum equipartition; Column~(7): maximum
  equipartition $\eta_{\rm max}$ reached, where $\sigma(m) \propto
  m^{-\eta}$. Column~(8): late-time asymptotic equipartition
  $\eta_{\infty}$ of the system as a whole. Columns~(5)--(7) pertain
  to the innermost $10\%$ of the stars as seen in projection, while
  column~(8) pertains to the system as a whole.

\smallskip

  Table notes. $^{*}$: canonical simulation discussed in
  Section~\ref{subsec:canon}; $\dag$: 30\% NS/BH retention; $\ddag$:
  $m_{\rm IMBH}=0.01 m_{\rm cluster}$; $\S$: ${r_t}=6.28$ [compact initial
    conditions where the Roche lobe is under-filled by a factor 2].  }
\end{minipage}
\end{table*}

\section{N-body Models}\label{sec:sim}

\subsection{Setup and Methodology}\label{subsec:simsetup}
   
We analyze the stellar dynamics in the direct N-body simulations
previously carried out by \citet{trenti10} to investigate the
evolution of structural parameters in star clusters. For full details
on the code and the simulation setup we refer to that paper (and
previous investigations in the same framework by \citealt{tre07a},
\citealt{gill08}, and \citealt{pas09}). To summarize, we follow the
dynamical evolution of simulated star clusters using the NBODY6 code
\citep{aar03}, which guarantees an exact treatment of multiple
interactions between stars by employing special regularization
techniques, without resorting to the introduction of softening. The
dynamics of the system are thus followed with extremely high accuracy,
at the price of a using lower number of particles compared to more
approximate methods (e.g., Monte Carlo codes).

We simulate systems with $N=32768$ to $N=65536$ particles, with an
initial mass function constructed to reproduce the long term evolution
of star clusters that are $\gtrsim 10$~Gyr old. As described in
\citet{trenti10}, we start from either a \citet{salp} or a \citet{ms}
initial mass function and then apply an instantaneous step of stellar
evolution with the \citet{hurley2000} evolutionary tracks to obtain a
main-sequence turnoff at $0.8~\mathrm{\msun}$. Stars in the range
$0.8~\mathrm{M_{\odot}}\leq m < 8.0~\mathrm{M_{\odot}}$ become white
dwarfs (with final masss prescribed by \citealt{hurley2000}), more
massive stars up to $25~\mathrm{M_{\odot}}$ become neutron stars,
while above $25~\mathrm{M_{\odot}}$ we form black holes in the
$5-10~\mathrm{M_{\odot}}$ mass range. Our standard assumption is to
retain $100\%$ of the dark remnants without assigning them velocity
kicks, but we explore a lower retention fraction ($30\%$) of neutron
stars and stellar mass black holes in two runs. During the simulation,
only gravitational interactions are considered and stars are not
evolved further. This simplified approach is ideal to highlight the
effects of gravitational forces on the long-term evolution of star
clusters and the drive toward energy equipartition, without the added
complication of disentangling dynamics from energy injection induced
by stellar evolution (through loss of mass with negative
energy). Stellar evolution affects star cluster structure
significantly only during the first few hundred Myr (e.g.  see
\citealt{hurley07,mackey08}), so in our analysis we are only
neglecting steady, low levels of mass loss induced by stellar
evolution at late times. These should not impact significantly the
dynamics, and indeed comparison between models with and without
stellar evolution does not show significant differences with respect
to predictions for mass segregation (see Figure~2 in
\citealt{gill08}). Based on that comparison, we expect that our
approach provides an approximate, yet accurate description of the
current evolution of globular clusters.

The initial conditions in the position and velocity space are drawn
from a King distribution function \citep{king66}, with scaled central
potential $W_0=3,5,7$. Initially, all particles of mass $m$ are drawn
from the same velocity distribution, hence at any given radius
$\sigma(m)$ is constant. Our runs include a primordial binary fraction
$f=N_b/(N_s+N_b) $ ranging between 0 and 0.1, with $N_s$ and $N_b$
being the number of singles and binaries respectively. All binaries
are ``hard'' in the \citet{heggie75} classification (i.e., semi-axis
typically smaller than $\sim 10$ AU for a typical globular cluster
with central velocity dispersion of $10$ km/s). One of our $N=32768$
runs contains a central IMBH, with the BH mass set at $1$ \% of the
total cluster mass. Table~\ref{tab:sim} gives an overview of all the
simulations performed and analyzed.

The simulated star clusters are tidally limited and particles
experience a tidal force from a point-like parent galaxy, assuming
that the cluster is in circular orbit at a distance selected so that
the tidal radius is self-consistently defined by the King density
profile, and all models fill their tidal radius initially (except for
one compact simulation listed in Table~\ref{tab:sim}). Hence, models
have different tidal field strengths for different $W_0$ values (see
\citealt{tre07a} and \citealt{trenti10} for a full definition of the
tidal field equations).

The simulations are run for $t\gtrsim 15~t_{rh}(0)$, where $t_{rh}(0)$
is the initial half-mass relaxation time, but they are stopped earlier
when $80\%$ or more of the initial mass in the system has been lost
due to evaporation of stars. In the natural (dimensionless) N-body
units defined by \citet{heg86} the relaxation time can by written as:
\be \label{trel}
  t_{rh} = \frac{0.138 N r_h^{3/2}}{log(0.11 N)},
\ee
where $r_h \approx 1$ is the half-mass radius and the time unit
approximately corresponds to the orbital period of a particle at
$r_h$. In N-body units, $t_{rh}(0)=497.7$ for the canonical simulation
highlighted in Table~\ref{tab:sim} ($N=65536$; $f=0$; $W_0=5$; Miller
\& Scalo IMF). For simplicity, all our time-dependent results are expressed in
units of this relaxation time, since this is the relevant timescale
for development of energy equipartition and thermodynamic equilibrium.

\subsection{Structural Evolution}\label{subsec:simstruc}

We refer to \citet{trenti10} for a detailed discussion of
the evolution of the cluster structural parameters. To summarize,
after a few relaxation times, all clusters tend to evolve toward a
quasi-universal core-halo structure with a central concentration
(measured either as the core to half mass radius ratio $r_c/r_h$ or as
core to tidal radius ratio $r_c/r_t$) which is independent of its
value at the beginning of the run. The long-term equilibrium value for
the concentration is defined uniquely by the efficiency of the central
energy production due to dynamical processes: runs with only single
particles undergo a marked core collapse and have higher
concentration, while in runs with primordial binaries the core
collapse is halted at lower central densities because of efficient
energy input by three body encounters. The run with an IMBH has the
largest core at late times. Once the long term $r_c/r_h$ ratio has
been established, the system undergoes a self-similar expansion at
fixed concentration, until the majority of the stars have been lost by
tidal evaporation, or the mechanism of central energy production is
depleted (for example because of disruption of all primordial
binaries).

This classical picture for the dynamical evolution of a globular
cluster is clearly present when one considers three-dimensional,
mass-based definitions of the core and half-mass radius
\citep{tre07a,fre07}. The situation becomes more complicated when
simulations are instead ``observed'', namely when only
two-dimensional, light-based definitions for structural parameters are
constructed \citep{trenti10}. For example, it is then possible that
clusters without primordial binaries still show a large core at late
times if the core is dominated by heavier dark remnants (neutron stars
and stellar mass black holes; \citealt{morscher13}).

\subsection{Analysis of Stellar Dynamics}\label{subsec:simdyn}

Our analysis here differs from that in \citet{trenti10} in that we now
consider the stellar dynamics in the simulations. Simulation snapshots
with particle positions and velocities were saved every $10$
dimensionless time units, yielding several tens of snapshots per
relaxation time. For each snapshot, at time $t$, we binned the
particles to derive the velocity dispersion $\sigma(m,r,t)$ as a
function of mass and projected radius. Mass binning was carried out in
intervals of $0.1~\mathrm{\msun}$. Two-dimensional (projected) radial
bins were defined by Lagrangian radii, each containing 10\% of the
particles in the snapshot. 

We took the following steps to minimize noise from low particle
numbers. First, for a given projection of the particle positions in
two dimensions, we measured the velocity dispersion along each of the
three independent cartesian coordinates, and then rescaled the results
to two-dimensional projection by averaging in quadrature and rescaling
to two-dimensions. This procedure is equivalent to measuring
three-dimensional velocity dispersions, and then rescaling the results
to two dimensions with a multiplying factor $\sqrt{2/3}$, and has been
done as a way to reduce shot noise in our results.\footnote{This
  procedure implicitly assumes isotropy in the three-dimensional
  velocity distribution, as expected for a collisionally relaxed
  stellar system. We did verify that our results do not change (except
  for somewhat increased noise) if we compute two-dimensional velocity
  dispersions directly.} Second, we created projections for the radial
distribution of the stars along three orthogonal directions and
averaged them. And third, we averaged together 30 independent
neighboring time snapshots. These steps allow us to achieve an
effective number of particle velocities in our analyses that for our
$N=65536$ runs is comparable to the number of stars in a massive
globular cluster like Omega Cen.

We determined the functional dependences of $\sigma(m,r,t)$ separately
for single main-sequence stars, binary stars, and compact remnants
(white dwarfs, neutron stars and stellar mass black holes) in the
simulations. For comparison to observational results, we focus in the
following primarily on the results for single main-sequence stars. In
reality, some stars in observational catalogs may be unresolved
binaries. However, we found that our main results described below did
not change significantly if binary stars in the simulations were
included in the single-star analyses (i.e., treating each binary of
total {\it luminosity} $L$ as a single star of that luminosity).

\section{Equipartition Results}\label{sec:results}

\subsection{Canonical Simulation}
\label{subsec:canon}

We describe first the results for a ``canonical'' simulation: a model
with $N=65536$ stars, a Miller \& Scalo initial mass function (IMF),
no primordial binaries, and a King concentration parameter $W_0=5$
(see Table~\ref{tab:sim}). This model has a moderate initial
concentration and external tidal field strength. As discussed in
\citet{trenti10}, the structural evolution of this cluster is fairly
proto-typical. Starting from an initial concentration
$c=-\log_{10}(r_t/r_c) \sim 1$, it undergoes core-collapse in about
$t\sim7 t_{rh}(0)$. At this stage, its observed concentration (based
on the core radius observed from the surface brightness profile) has
reached a quasi-equilibrium value $c\sim 1.7$. The three-dimensional
concentration is instead higher ($c\gtrsim 2.5$) as the 3D core
contracted to reach densities sufficient to produce a few binaries
through three body encounters. At $t\gtrsim 7t_{rh}(0)$, the system
continues to evolve with self-similar density and surface-brightness
profiles, progressively losing mass as a result of the external tidal
field until complete dissolution at $t\sim 16 t_{rh}(0)$.

\subsubsection{Mass Dependence of Kinematics}
\label{subsec:canoneta}

\begin{figure}
\resizebox{\hsize}{!}{\includegraphics{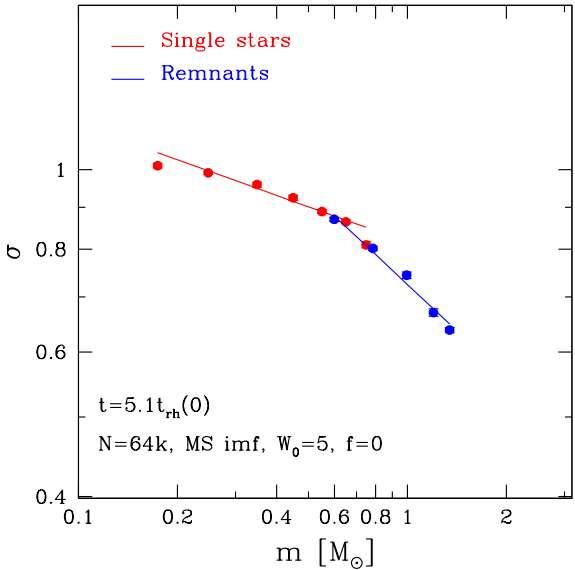}}
\caption{Velocity dispersion $\sigma$ as a function of particle mass
  $m$ around time $t=5.1 t_{rh}(0)$ for particles at the center of the
  system (two-dimensional projected radius $r<0.78~r_c$) in our
  canonical $N$-body simulation (the $N=65536$ run with $W_0=5$, a
  \citet{ms} IMF, and no primordial binaries). The red points with
  errorbars are for single main sequence stars, while the blue points
  are for compact remnants. Lines show the best fitting power-law
  relation $\sigma(m) \propto m^{-\eta}$ for each set of objects, with
  $\eta = 0.14$ for single main sequence stars and $\eta = 0.34$ for
  compact remnants.}\label{fig:sig_M}
\end{figure}

Figure~\ref{fig:sig_M} shows $\sigma(m)$, determined as described in
Section~\ref{subsec:simdyn}, for the innermost radial Lagrangian bin
($10\%$ enclosed mass) at $t=5.1~t_{rh}(0)$. This is during the
initial core contraction as the system is on its way toward core
collapse. Red points pertain to single stars along the main sequence,
while blue points pertain to compact remnants which are generally
heavier. Single stars are found up to the turnoff mass of $0.8
~\mathrm{\msun}$, while the lightest compact objects have masses of
$\sim 0.55~\mathrm{\msun}$.

\begin{figure}
\resizebox{\hsize}{!}{\includegraphics{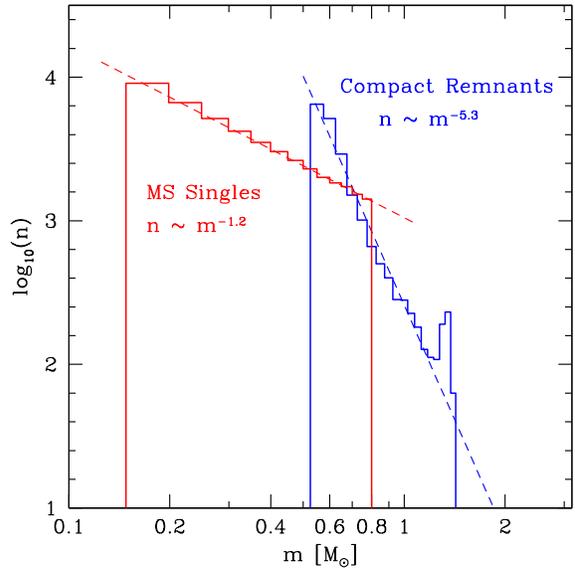}}
\caption{Mass function for the particles of an $N$-body run
  (including, e.g., our canonical simulation) starting from a
  \citealt{ms} IMF, evolved to a turn-off mass of
  $0.8~\mathrm{\msun}$. The red histogram represents main-sequence
  stars, and the blue one compact remnants. Approximate power-law fits
  are show as dashed curves, which have slopes $m^{-1.2}$ and
  $m^{-5.3}$, respectively. Since the mass function for remnants is
  much steeper, it is easier for them to approach energy equipartition
  (see Figure~\ref{fig:sig_M}).}\label{fig:initial_mass}
\end{figure}

The main sequence stars and the compact remnants have comparable
dispersions over the mass range where both types of objects exist,
$0.55 \lesssim m \lesssim 0.8~\mathrm{\msun}$. However, when the full
range is considered over which each object type is found, the compact
remnants have a significantly steeper $\sigma(m)$ relation than the
main-sequence stars (approximately $m^{-0.34}$ vs.~$m^{-0.14}$,
respectively). Neither type of object achieves energy equipartition,
but the heavier compact objects are closer to it than the lighter
main-sequence stars. This behavior is qualitatively consistent with
the stability analysis presented by \citet{vishniac78}, given that the
remnants have a significantly steeper mass function than the
main-sequence stars (see Figure~\ref{fig:initial_mass}). This makes it
easier for the population in the $0.7-2~\mathrm{\msun}$ mass range to
thermalize. The steep mass function for remnants is due to the
significant mass loss experienced by massive stars before they become
remnants, and is relatively insensitive to the adopted IMF. The
generic shape of $\sigma(m)$ in Figure~\ref{fig:sig_M} is therefore
typical for all our simulations.

The relation $\sigma(m)$ in Figure~\ref{fig:sig_M} shows what appears
to be a break centered around $m \approx 0.7~\mathrm{\msun}$, with some
additional curvature towards lower
masses. Figure~\ref{fig:initial_mass} shows that $m \approx 0.7~\mathrm{\msun}$
is also the mass at which the density of main-sequence stars is
equal to the density of the remnants, so this might provide some clues
to the origin of the break\footnote{This is obtained considering all
  particles in the system. In general the distributions change with
  radius because main sequence stars and remnants have different
  mass-segregation profiles. However, the intersection point of the
  two curves does not depend on radius, since by definition particles
  at the intersection point of the two curves have the same mass in
  both samples, and hence the same mass segregation profile.}. The
\cite{vishniac78} analysis suggests that more energy equipartition is
possible for a sub-population within a given mass range, when the
lighter counterparts are much more abundant. So thermalization is more
likely for objects above $0.7~\mathrm{\msun}$, as there are plenty of
lower mass particles along a steep mass function ($m^{-5.3}$). By
contrast, the lowest mass remnants and the main sequence stars below
$0.7~\mathrm{\msun}$ are less able to thermalize efficiently, because
the mass function is only increasing mildly ($m^{-1.2}$). In
conclusion, this might explain the break in Figure~\ref{fig:sig_M}.

Figure~\ref{fig:sig_M} shows that the $\sigma(m)$ relations for single
main-sequence stars and compact remnants separately, are both
reasonably well fit by a power law of the form $\sigma(m) \propto
m^{-\eta}$. The compact remnants are only of theoretical interest as
they are not directly observable. In the following, we therefore focus
our attention only on the main sequence stars, which can be
observed. Using the $\sigma(m,r,t)$ profiles constructed as described
in Section~\ref{subsec:simdyn}, we fit for each radial Lagrangian bin
$r$ and at each time $t$ the mass-dependent kinematics assuming a
power law $\sigma(m) \propto m^{-\eta}$ over the mass range $0.2\leq
m/\mathrm{\msun} \leq 0.7$. Here $\eta$ is a free parameter, with
$\eta=0$ corresponding to absence of equipartition (velocity
dispersion independent of mass), $0 < \eta < 0.5$ corresponding to
partial equipartition, and $\eta = 0.5$ corresponding to complete
equipartition. The quantity $\eta$ is a well-defined fit-parameter
that can be compared across models, even when the $\sigma(m)$ relation
shows some residual curvature, as e.g.~in Figure~\ref{fig:sig_M}.

\subsubsection{Radial and Time Dependence of Equipartition}
\label{subsec:canondep}

\begin{figure}
\resizebox{\hsize}{!}{\includegraphics{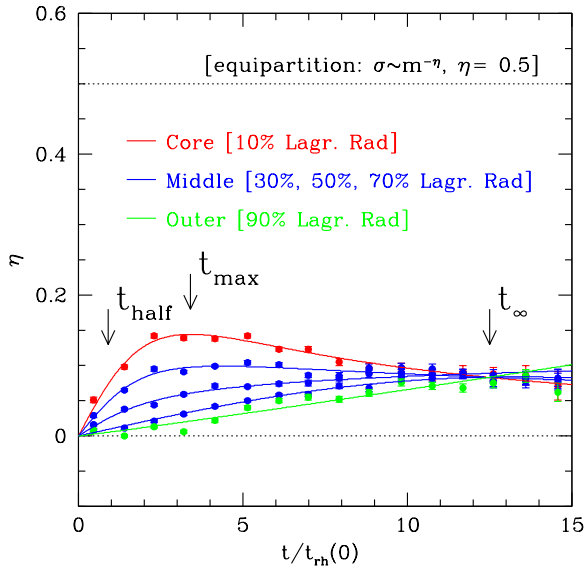}}
\caption{Time evolution of the energy equipartition indicator $\eta$
  (the power-law slope in the relation between velocity dispersion and
  mass, $\sigma \propto m^{-\eta}$) for single main sequence stars in
  our canonical $N$-body simulation. Data points with error bars were
  determined directly from the simulation snapshots, as described in
  Sections~\ref{subsec:simdyn} and~\ref{subsec:canoneta}. Solid curves
  are fits of the form given by equation~(\ref{eq:etafit}). Red refers
  to the innermost $10\%$ of the main sequence stars in projection;
  blue refers to three projected annuli containing $10\%$ of the
  particles each, and ending at $30\%$, $50\%$ and $70\%$ of the
  enclosed particle number; green refers to the outer Lagrangian bin
  at $90\%$ of the enclosed particle number. The time along the
  abscissa is expressed in units of the initial half-mass relaxation
  time $t_{rh}(0)$. These results for the canonical simulation are
  representative of the general behavior of our $N$-body
  simulations. The inner region of the system reaches a higher degree
  of equipartition within a few relaxation times than the outer
  regions. An increase in $\eta$ takes longer to develop at large
  radii. However, at later times, the whole system converges toward a
  common value of $\eta$. Arrows indicate the characteristic times
  defined in Section~\ref{subsec:canondep}. Complete energy
  equipartition ($\eta=0.5$) is never attained, confirming previous
  investigations based on stability analysis.}\label{fig:eta_evol}
\end{figure}

The degree of energy equipartition as a function of time, $\eta(t)$,
is shown for different Lagrangian radii in the canonical simulation in
Figure~\ref{fig:eta_evol}. At $t=0$, the system starts (by
construction) with no equipartition at all. As time progresses, there
is a relatively rapid linear rise in $\eta(t)$ for the inner
Lagrangian radii. This is followed by a flattening toward a maximum
value around $\eta \approx 0.15$ that is reached after a few initial
half-mass relaxation times. Subsequently, the value of $\eta$ drops
slowly\footnote{This drop in $\eta$ is likely due to formation of
  binaries in the core. These provide heating that acts to reduce mass
  segregation and the approach to equipartition.}. For the outer
Lagrangian radii, the value of $\eta(t)$ evolves on a longer
timescale, reflecting the longer relaxation time in the outskirts of
the system. For {\it all} radii there is convergence to a value around
$\eta \approx 0.10$ at late times.  However, because of the slower
development of equipartition, the outer regions do not first
``overshoot'' this value by reaching a maximum at early times, as seen
for the inner Lagrangian radii.

To describe the time evolution of equipartition empirically, we
adopted a fit function of the form:
\be\label{eq:etafit}
  \eta_{\rm fit}(t,r) = \eta_c(r) \times f_a[t/t_c(r)],
\ee
where
\be
f_a[x] = x \times {(1+x^2)^{[a(r)-1]/2}} .
\ee
Here $t_c(r)$ and $\eta_c(r)$ are a characteristic time and a
characteristic $\eta$ value, both of which depend on radius. The
parameter $a(r)$ also depends on radius and defines the exact shape of
the function $f_a$. Upon varying $t_c$, $\eta_c$ and $a$ for each
radial bin to optimize the fit, this empirical formula generally
provides an adequate description of the time dependence of
equipartition in the simulations. For the canonical simulation, this
is illustrated by the solid curves in Figure~\ref{fig:eta_evol}.

To compare the results of different simulations, it proved convenient
to extract the following quantities, which capture the essence of the
time evolution of equipartition. The quantities were determined for
all simulations from the empirical fits of the form given by
equation~\ref{eq:etafit}, rather than from the individual simulation
snapshot datapoints, because the latter are available only at discrete
times.
  
\begin{itemize} 

\item $\eta_{\rm max}$: the maximum $\eta$ that is reached in the
  central region, defined here as the innermost Lagrangian radius
  (inner 10\% of the projected mass).

\smallskip

\item $t_{\rm max}$: the time at which the central region achieves
  $\eta = \eta_{\rm max}$.

\smallskip

\item $t_{\rm half}$: the time at which the central region achieves 
  $\eta = 0.5 \eta_{\rm max}$. This value is defined more robustly than
  $t_{\rm max}$, since $\eta(t)$ tends to flatten near its maximum. 

\smallskip

\item $t_{\infty}$: the late time $t > t_{\rm max}$ at which the
  $\eta$ values for all Lagrangian radii are most similar (defined as
  minimum dispersion in $\eta$).
 
\smallskip

\item $\eta_{\infty}$: the value of $\eta$ at $t = t_{\infty}$
  averaged over all Lagrangian bins. This represents approximately the
  long-term asymptotic value of $\eta$ for the system as a whole.

\end{itemize}

\noindent We list these quantities in Table~\ref{tab:sim} for all
simulations. We omit $t_{\infty}$ since it is not very robustly
determined, lacks a clear physical meaning, and may depend on the
(arbitrary) time at which the simulation is ended. Since $\eta(t)$ is
roughly flat at large times, $\eta_{\infty}$ by contrast is defined
more robustly.

For the canonical simulation, $t_{\rm half} = 0.9 t_{rh}(0)$, $t_{\rm
  max} = 3.4 t_{rh}(0)$, and $t_{\infty} = 12.5 t_{rh}(0)$. These
times are indicated with arrows in Figure~\ref{fig:eta_evol}. Half of
the maximum equipartition develops rather quickly in the central
region of the system, while the maximum value is reached only around
half-way to core-collapse. The maximum equipartition is characterized
by $\eta_{\rm max} = 0.144 \pm 0.006$, and the long-term asymptotic
value is $\eta_{\rm \infty} = 0.085 \pm 0.006$.

The analysis discussed above was carried out using projected radii,
which are the only observable quantities. The results show that the
canonical simulation never attains complete energy
equipartition. However, at early times the particles in the projected
core are closer to complete equipartition than what is found for all
particles at late times. So it expected that stars in the {\it
  three-dimensional} core might achieve an even higher degree of
equipartition at early times. We analyzed $\eta$ also for the three
dimensional core, and found that this is indeed the case. However the
effect is very modest, and the three-dimensional core is only
marginally closer to equipartition (by $\Delta \eta \approx 0.05$ at
$t=t_{\rm max}$). Therefore, no region in the simulation is attaining
complete energy equipartition, in full agreement with the analytical
stability analysis of \citet{vishniac78}.

\begin{figure}
\resizebox{\hsize}{!}{\includegraphics{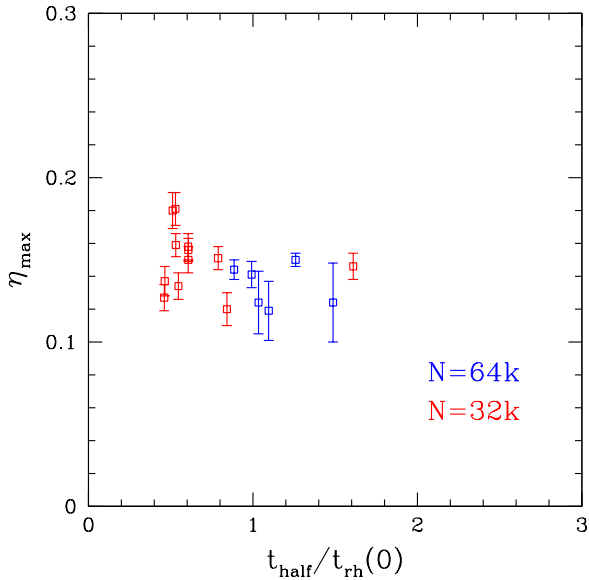}}
\caption{Scatter plot of equipartition results, showing $\eta_{\rm
    max}$ vs.~$t_{\rm half}$ for all of the $N$-body simulations. The
  velocity dispersion of single main-sequence stars scales as $\sigma
  \propto m^{-\eta}$, and the quantity $\eta_{\rm max}$ is the maximum
  power-law slope $\eta$ for the innermost $10\%$ radial Lagrangian
  bin in projection. The quantity $t_{\rm half}$ is the time at which
  this radial bin reaches $\eta = 0.5 \eta_{\rm max}$. Simulations
  with $N=32k$ particles are shown in red, and those with $N=64k$
  particles in blue. The initial rise in $\eta(t)$ is approximately
  linear, so all simulations have essentially achieved a near-maximum
  equipartition in their cores after $t \gtrsim 3~t_{rh}(0)$.}\label{fig:eta_half}
\end{figure}

\begin{figure}
\resizebox{\hsize}{!}{\includegraphics{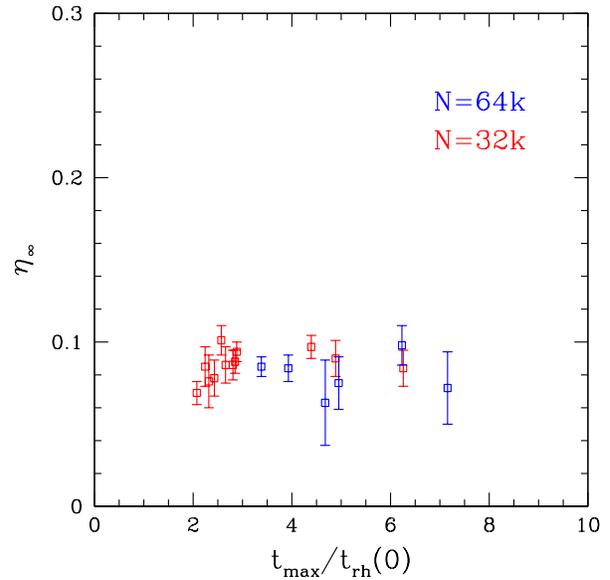}}
\caption{Scatter plot of equipartition results, showing
  $\eta_{\infty}$ vs.~$t_{\rm max}$ for all of the $N$-body
  simulations. The velocity dispersion of single main-sequence stars
  scales as $\sigma \propto m^{-\eta}$, and the quantity
  $\eta_{\infty}$ is the asymptotic power-law slope toward which the
  system as a whole converges at late times. The quantity $t_{\rm
    max}$ is the time at which the innermost $10\%$ radial Lagrangian
  bin in projection achieves its maximum value $\eta_{\rm max}$.
  Simulations with $N=32k$ particles are shown in red, and those with
  $N=64k$ particles in blue. The quantity $\eta_{\infty}$ has a
  near-universal value of $0.08 \pm 0.02$ across all simulations, and
  is very far from complete equipartition ($\eta =
  0.5$).}\label{fig:eta_inf}
\end{figure}

\subsection{Dependence on Model Parameters}

\subsubsection{Sample Statistics}
\label{subsec:sampstat}

We have found that the analysis of energy equipartition across our
sample of $N$-body simulations shows a surprisingly uniform and
consistent picture. All the runs behave to first approximation like
the canonical run discussed in Section~\ref{subsec:canon}.
Figures~\ref{fig:eta_half} and~\ref{fig:eta_inf} show scatter-plot
representations the results for $t_{\rm half}$, $t_{\rm max}$,
$\eta_{\rm max}$ and $\eta_{\infty}$ for all of the simulations. The
uncertainties in $\eta_{max}$ and $\eta_{\infty}$ are generally small,
$\Delta \eta \lesssim 0.01$, and are smaller than the scatter between
different simulations.

The value of $\eta_{\rm max}$ varies from $\eta_{max}=0.12$ achieved
by the run with a central IMBH and by a run with low concentration
($W_0=3$) and primordial binaries ($f=0.055$) to $\eta_{max}=0.18$ for
the runs with a low retention fraction of neutron stars and stellar
mass black holes. Runs with a higher fraction of primordial binaries
develop less equipartition than their similar counterparts with single
stars only, although the effect is overall modest (e.g. from
$\eta_{max}=0.159$ to $\eta_{max}=0.127$ going from $f=0$ to $f=0.1$
for the series of runs with $N=32768$ particles, $W_0=7$ and a
\citet{ms} IMF). Comparing runs with a \citet{salp} versus \citet{ms}
IMF does not show any significant difference (we are neglecting
stellar evolution which could change somewhat this conclusion, but we
discuss in Section~\ref{subsec:simsetup} how we expect such impact to
be modest at most). Table~\ref{tab:sim} also shows that lower initial
concentration $W_0$ generally yields slightly lower $\eta_{max}$. As
discussed in Section~\ref{subsec:interp} below, this is most likely
due to the different tidal field strength assumed in simulations with
different $W_0$, rather than due to the concentration itself.

The long-term asymptotic value $\eta_{\infty}$ is even more uniform
across the sample of simulations than is $\eta_{\rm max}$. Its value
ranges from $\eta_{\infty}=0.063$ to $\eta_{\infty}=0.101$. The
sensitivity of $\eta_{\infty}$ to model parameters is qualitatively
similar to that for $\eta_{\rm max}$, and there is a mild correlation
between the two quantities across the sample. A Spearman rank
correlation test gives a correlation $r=0.36$, which is significant at
greater than $90\%$ confidence for our sample of 18 simulations. Since
some of the mechanisms that drive the suppression of the equipartition
are most active in the core rather than in the whole system, such as
the presence of an IMBH, it is not surprising to observe only a modest
correlation.

\begin{figure}
\resizebox{\hsize}{!}{\includegraphics{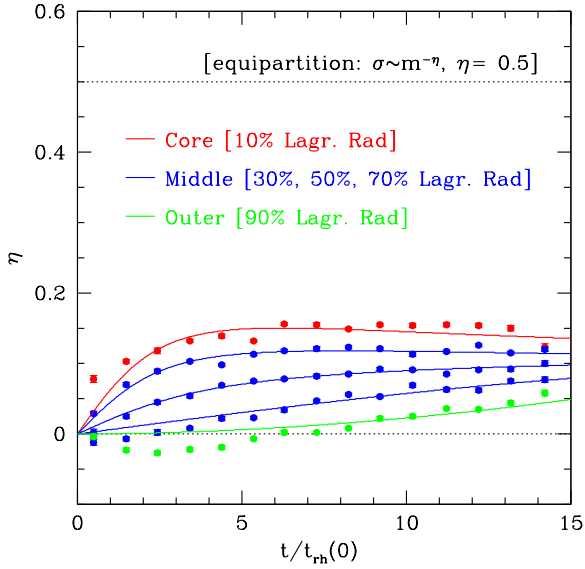}}
\caption{Time evolution of the energy equipartition indicator $\eta$
  (the power-law slope in the relation between velocity dispersion and
  mass, $\sigma \propto m^{-\eta}$) for single main sequence stars in
  a simulation with $N=65536$ particles, $W_0=7$, a \citet{salp} IMF,
  and no primordial binaries. The overall evolution is similar to that
  for the canonical simulation (see Figure~\ref{fig:eta_evol}, which
  uses the same layout and color-coding). However, the present
  simulation shows ``energy divergence'' in the outer parts at early
  times, with evolution toward $\eta < 0$ (i.e., high-mass stars have
  higher velocity dispersion than low-mass stars). This transient
  behavior is possibly related to ejection of particles from the core
  into the outer parts of the system because of the dynamical
  interactions (see Section~\ref{subsec:diverge}). This simulation
  also shows that the best-fit function of the form given by
  equation~(\ref{eq:etafit}) may underestimate a very rapid rise in
  $\eta(t)$ at early times, yielding a value of $t_{\rm half}$ that is
  biased high.}\label{fig:eta_time_64k_vanilla}
\end{figure}

The timescale needed to approach maximum equipartition, as measured by
either $t_{\rm half}$ or $t_{\rm max}$, varies more between
simulations. The value of $t_{\rm half}/t_{rh}(0)$ ranges from
approximately $0.5$--$1.6$. The initial rise in $\eta(t)$ is
approximately linear, so this implies that after $t \gtrsim
3~t_{rh}(0)$ all systems have essentially achieved a near-maximum
equipartition in their cores. The actual values of $t_{\rm
  max}/t_{rh}(0)$ show considerable variation though, ranging from
$2.1$--$7.2$. This is because the $\eta(t)$ profile beyond the
maximum is often nearly flat for the innermost Lagrangian bin, so that
$t_{\rm max}$ is not robustly determined. An example of this is
provided in Figure~\ref{fig:eta_time_64k_vanilla}, which is similar to
our canonical simulation in Section~\ref{subsec:canon}, but pertains
to an initial King concentration index $W_0 = 7$.

Figure~\ref{fig:eta_half} shows that the runs with $N=65536$ particles
tend to have longer $t_{\rm half}/t_{rh}(0)$ than those with
$N=32768$. However, this may be due to small systematic errors in the
determination of $t_{\rm half}$. Figure~\ref{fig:eta_time_64k_vanilla}
shows that the steep initial rise in $\eta(t)$ in the inner Lagrangian
bin is not always well reproduced by the fit of the functional form in
equation~(\ref{eq:etafit}). The values of $\eta_{\rm max}$,
$\eta_{\infty}$ and $t_{\rm max}$ do not show any clear trends with
particle number. This suggests that our simulations have sufficient
$N$ to determine real physical trends (as opposed to numerical
artifacts, transients, or shot-noise dominated features), and that any
dependence on $N$ is properly accounted for by expressing time scales
in units of $t_{rh}(0)$.

The equipartition timescales $t_{\rm half}$ and $t_{\rm max}$ do not
show obvious correlations with other parameters or with the model
initial conditions. There is a hint of negative correlation between
$t_{\rm half}$ and $\eta_{\rm max}$ (Figure~\ref{fig:eta_half}), but
we are hesitant to attach much significance to this, given the
possibility of small systematic errors in the determination of $t_{\rm
  half}$. No obvious correlation is present between $t_{\rm max}$ and
$\eta_{\infty}$ (Figure~\ref{fig:eta_inf}).

\begin{figure}
\resizebox{\hsize}{!}{\includegraphics{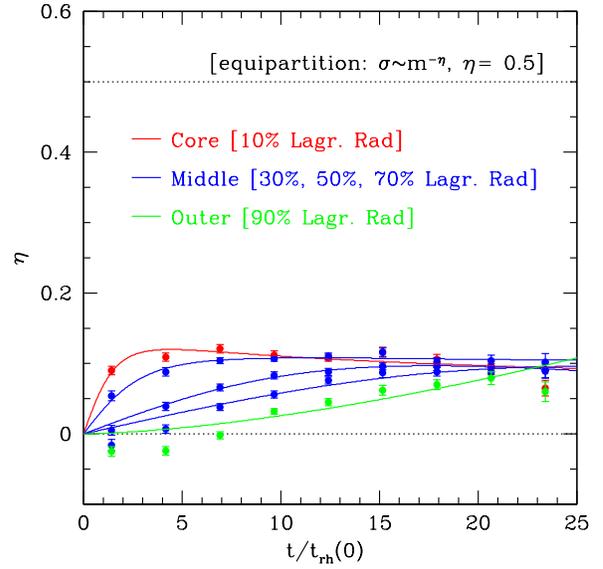}}
\caption{Time evolution of the energy equipartition indicator $\eta$
  (the power-law slope in the relation between velocity dispersion and
  mass, $\sigma \propto m^{-\eta}$) for single main sequence stars in
  a simulation with a central IMBH, $N=32769$ particles, $W_0=7$, a
  \citet{ms} IMF, and no primordial binaries. The overall evolution is
  similar to that for the corresponding simulation without a central
  IMBH (see Figure~\ref{fig:eta_time_64k_vanilla}, which uses the same
  layout and color-coding). However, the IMBH suppresses the amount of
  equipartition that is achieved (see Section~\ref{subsec:interp}),
  yielding somewhat lower $\eta$, especially for the innermost radial
  Lagrangian bin. This simulation also shows the same ``energy
  divergence'' toward $\eta < 0$ in the outer parts at early times as
  in Figure~\ref{fig:eta_time_64k_vanilla}.}\label{fig:eta_evol_IMBH}
\end{figure}

\subsubsection{Interpretation }
\label{subsec:interp}

The maximum equipartition value $\eta_{\rm max}$ correlates with model
initial conditions in a way that can be understood based on the
physical insights, stability analyses, and mass segregation research
discussed in Section~\ref{sec:intro}.

It has been shown previously that the presence of an IMBH suppresses
mass segregation. So the effect is naturally to reduce the amount of
energy equipartition, measured by $\eta$, as
well. Figure~\ref{fig:eta_evol_IMBH} shows the $\eta(t,r)$ profiles
for our simulation with an IMBH. These can be compared to the results
in Figure~\ref{fig:eta_time_64k_vanilla} for a similar model (with
twice the number of particles) without an IMBH. The inclusion of an
IMBH yields a modest reduction in $\eta_{\rm max}$. We attribute this
to the fact that the IMBH generally has at least one particle tightly
bound to it (often a compact remnant). This binary system scatters
single main sequence stars out of the core, independently of their
mass (as these third bodies have masses much smaller than the IMBH
binary). This partially counters the natural tendency for more massive
main-sequence stars to segregate into the core and have a lower
velocity dispersion compared to lighter counterparts.

Primordial binaries also act toward suppressing mass segregation, with
a qualitatively similar mechanism. So it is not surprisingly that we
generally find a decrease in $\eta_{max}$ with increasing $f$ (see
Table~\ref{tab:sim}). And stellar mass black holes tend to form BH-BH
binaries, which also act in a similar way. Hence, it is not surprising
that our two simulations with a {\it low} retention fraction of
neutron stars and BHs, yield the highest values of $\eta_{\rm  max}$
(i.e., achieving the most equipartition in their central region).

Finally, we find that runs with a steeper IMF yield similar $\eta_{\rm
  max}$ but marginally higher $\eta_{\infty}$. This is consistent with
the analysis of \citet{vishniac78}. With a steeper IMF, heavy
particles can transfer their kinetic energy more efficiently to their
lighter counterparts, since those are more numerous. Hence, more
equipartition can be reached.

The theoretical expectation is that the initial concentration of the
system should not play a significant role in determining $\eta_{\rm
  max}$.  This is because after several relaxation times the system
has evolved toward a quasi-equilibrium state with a universal
concentration, independently from where it started from
\citep{trenti10}. As mentioned in Section~\ref{subsec:sampstat}, our
results taken at face value do not confirm this. We find lower
$\eta_{\rm max}$ in simulations started with lower $W_0$. However,
this is most likely because of the effects of the tidal field on the
kinematics of the system. The concentration in the initial conditions
of most of our simulations is self-consistently associated with the
assumed tidal field strength. Instead, our simulation starting from
low initial concentration $W_0=3$, but with an under-filled tidal
Roche lobe, yields similar $\eta_{\rm max}$ as a simulation starting
with $W_0=7$. So we conclude tentatively that concentration does not
affect equipartition, while a strong tidal field has a mild effect of
suppressing the maximum $\eta$.

\subsection{Energy Divergence}
\label{subsec:diverge}

The expectation based on generic thermodynamic arguments is that
clusters evolve toward energy equipartition (even if they may never
reach complete equipartition). So if one stars from a state with
$\eta=0$ (velocity dispersion independent of mass $m$), one expects
evolution toward a state in which high-mass stars have lower velocity
dispersions than low-mass stars (i.e., $\eta > 0$). Interestingly, we
have found a few instances in our simulations where this does not hold
true. Instead, there is what one might call ``energy divergence'',
where part of system evolves toward a state where high-mass stars have
{\it higher} velocity dispersions than low-mass stars (i.e., $\eta <
0$).

We find such energy divergence only in some of our simulations, and
only for the outer $\sim 20-30\%$ of the particles during the first
few initial half-mass relaxation times. The $\eta(t,r)$ profiles shown
in both Figures~\ref{fig:eta_time_64k_vanilla}
and~\ref{fig:eta_evol_IMBH} provide examples of this. The energy
divergence appears transient, and it develops over the same timescale
over which maximum energy equipartition develops in the core. Because
of this, we interpret the energy divergence as a result of dynamical
interactions in the core.  These interactions eject to the outer
regions of the system particles that have partially thermalized at a
higher velocity dispersion, and that tend to be heavier than those
native in the outer regions because of central mass segregation.

Support for this scenario comes from the fact that the two simulations
that clearly show this feature are the IMBH run
(Figure~\ref{fig:eta_evol_IMBH}) and a run that forms a BH-BH binary
and starts from a high core density ($W_0=7$;
Figure~\ref{fig:eta_time_64k_vanilla}). In both cases, it is expected
that main-sequence stars in the core are scattered out nearly
independently of their mass \citep{gill08}. After several half-mass
relaxation times, this transient disappears once partial equipartition
has time to develop in the outer regions as well.

We also find $\eta < 0$ in some of our simulations at very late times,
when the system has lost more than $80\%$ of its mass. However, this
may very well be because very few main-sequence stars are remaining,
so that the measurement of $\eta$ suffers from significant numerical
noise. We therefore do not attach much physical significance to those
instances of energy divergence.

\section{Application to Omega Cen}
\label{sec:omegacen}

With the advent of HST proper motion studies of the internal
kinematics in globular clusters (see Section~\ref{sec:intro}), it is
now becoming possible to actually {\it measure} the amount of
equipartition in globular clusters. With the models presented in this
paper, this opens up the possibility of detailed data-model
comparisons. As a first application, we consider here the case of the
galactic globular cluster Omega Cen (NGC 5139).

\citet{anderson} determined a proper motion catalog for the core of
Omega Cen from ACS/WFC data with a 4-year time baseline. The relation
between $\sigma$ and $m$ for stars along the main sequence with $m =
0.5$--$0.8\msun$ is shown in their Figure~25c. They showed that a
power-law $\sigma = m^{-\eta}$ with $\eta \approx 0.2$ describes the
data reasonably well.  This implies that Omega Cen is between the
extremes of no equipartition and complete equipartition. This is
qualitatively similar to what we find generically in our $N$-body
simulations. It is also consistent with earlier work by
\citet{anderson02}, which showed that there is mass segregation in
Omega Cen, but not as much as predicted by multi-mass Michie-King
models that assume complete equipartition.

\begin{figure}
\resizebox{\hsize}{!}{\includegraphics{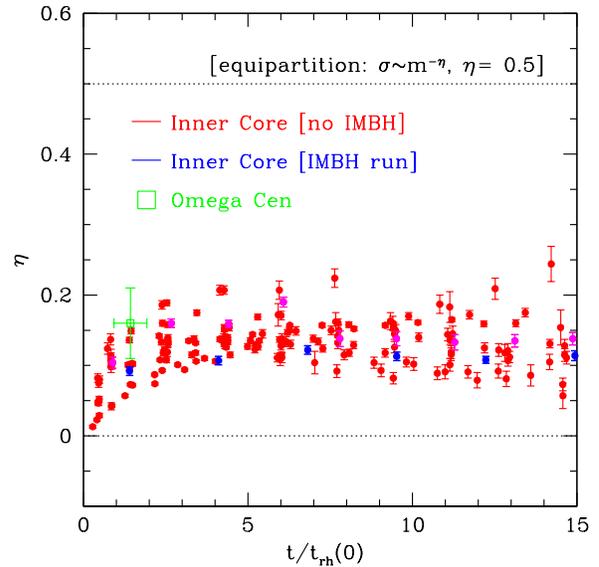}}
\caption{Data-model comparison of energy equipartition in Omega Cen.
  The time evolution of $\eta$ (the power-law slope in the relation
  between velocity dispersion and mass, $\sigma \propto m^{-\eta}$) is
  shown as in Figures~\ref{fig:eta_evol},
  \ref{fig:eta_time_64k_vanilla} and~\ref{fig:eta_evol_IMBH}. Data
  points show the results from the $N$-body simulation snapshots for
  the innermost $10\%$ of the main sequence stars in projection. All
  $N$-body simulations are included, and are generally shown in red;
  the simulation that includes an IMBH is shown in blue, and the
  simulation that is similar to that but without the IMBH is shown in
  magenta. The observed $\eta$ for Omega Cen is plotted with error
  bars in green at the current dynamical age $t/t_{rh}(0)$. Overall,
  there is excellent consistency between simulations and observations,
  supporting the view that globular clusters are not generally in
  energy equipartition.}\label{fig:omega_comparison}
\end{figure}

Andrea Bellini (priv.~comm.~2013) refined the proper motion catalog of
\citet{anderson} in the context of an ongoing HST Archival study of
two dozen globular clusters \citep{bellini}. He added more recent
WFC3/UVIS data in many filters. This increases the time-baseline of
the Omega Cen data to 8 years, and reduces all proper motion errors
correspondingly. For the same sample of stars as in the ``central
field'' of \citet{anderson}, including stars down to $m \approx
0.3\msun$ but using improved proper motions, he finds that $\eta = 0.16 \pm 0.05$. This is the value
that we will adopt here for our data-model comparison. The error bar
includes the contributions of both random and systematic
errors. Random errors are obtained from a least-squares fit to the
$\sigma(m)$ data, with propagation of the random uncertainties for
individual mass bins. Systematic errors reflect, e.g., variations in
best-fit $\eta$ values depending on exactly which WFC3/UVIS filters
are included in the construction of the catalog.

For a quantitative data-model comparison, we need to estimate the
dynamical age of Omega Cen. \citet{mcm05} showed that the surface
brightness is well fitted by a King model with $W_0=6.2^{+0.2}_{-0.1}$,
$c=1.31^{+0.05}_{-0.03}$, $r_c=141.''2^{+6.''5}_{-12.''69}$, and
$L_{\rm V,tot} = 10.0^{6.10 \pm 0.02}$. \citet{vdm10} derived from a
dynamical model for the spatially resolved kinematics a distance
$D=4.7\pm0.06$ kpc and mass-to-light ratio $M/L_V = 2.64 \pm 0.03$ in
solar units. From these values we infer using equation~(\ref{trel}) a
relaxation time $t_{rh} = 10^{9.96}$ yr at the half-light radius $r_h
= 6.8$ pc. For an assumed age of $13$ Gyr, this implies
$t=1.43~t_{rh}$. Note that our rescaling of the simulation results to
the relaxation time age of Omega Cen can be done because we are
neglecting stellar evolution. Otherwise, rescaling the dynamical times
would have modified the physical age of the stars, changing the
dynamics (for example by having a different turn-off mass). 

For comparison to our simulations, we need to know the age in units of
the {\it initial} half-mass relaxation time $t_{rh}(0)$, and not the
{\it current} half-mass relaxation time $t_{rh}$. Because of a
combination of tidal evaporation of particles and
contraction/expansion of the system, there is not a well defined
relation between current and initial relaxation time. In the
case of a system similar to Omega Cen, it is possible that $t_{rh}(t)$
is either larger or smaller than $t_{rh}(0)$, primarily depending on
the retention fraction of stellar mass black holes (and on the
presence/absence of a central IMBH), and on the initial concentration
of the system. If either stellar mass black holes or an IMBH are
present in the system, then likely there has been expansion of the
half-light radius fueled by energy generation in the core of the
system, which implies $t_{rh}(t)>t_{rh}(0)$. On the other hand, if
Omega Cen lacks massive dark remnants, it is possible that its
half-light radius has undergone contraction since the cluster
formation. Conservatively, we assign an uncertainty of $35\%$ to the
dynamic age of the system: $t=(1.43 \pm 0.5)~t_{rh}(0)$. This
uncertainty includes both contributions propagated from the current
model parameters that describe Omega Cen, and from the systematic
unknowns on the composition and expansion/contraction history of the
cluster.

To compare the observed amount of energy equipartition in the central
field of \cite{anderson} to the simulations, we use the simulated
Lagrangian bin that contains the inner 10\% of the particles in
projection. When the simulation is scaled to the size of Omega Cen,
these particles have a similar median projected radius as the observed
field.\footnote{This comparison does not account for all the details
  of the observations and of the cluster. For example, the field
  observed for proper motions is not a circle. Omega Cen is elongated
  and rotating, unlike the simulated clusters. And finally, the mass
  function in our simulations was not tailored to fit specific
  observations of Omega Cen.} The comparison between the measured
$\eta$ for Omega Cen and the predictions of {\it all} of the $N$-body
simulations is shown in
Figure~\ref{fig:omega_comparison}.\footnote{Most of our simulations
  have initial concentration $W_0 = 5$ or $7$, which is close to the
  current value $W_0=6.2^{+0.2}_{-0.1}$ for Omega Cen.}

Given the dynamical age of Omega Cen, the simulations predict $\eta$
to be in the range $0.07$--$0.15$. The measured $\eta = 0.16 \pm 0.05$
falls within this range. So overall, there is excellent consistency
between simulations and observations. The simulation with an IMBH
predicts a slightly lower $\eta$ than observed, while the simulation
with the same initial conditions but no IMBH fits better. However, a
broader suite of simulations with IMBHs would be needed to place any
quantitative constraints on the possible presence of an IMBH in Omega
Cen (a topic that continues to be debated, e.g.,
\citet{noyola10},\citet{vdm10}). With more sophisticated future
analyses, smaller observational uncertainties, or larger samples of
clusters, it may become possible to use observed $\eta$ values to
discriminate between different clusters models. 

\section{Discussion and Conclusions}\label{sec:con}

It is widely believed, and commonly taught, that a globular cluster
evolves, given a sufficiently long time, toward a state in which its
stars are in energy equipartition (at least near the center, where the
relaxation times are shortest). If the mean kinetic energy $\langle
{1\over2}m v^2\rangle$ becomes independent of mass $m$ due to two-body
relaxation (i.e., collisions), the velocity dispersion $\sigma \equiv
\langle v^2\rangle^{0.5}$ scales as $\sigma \propto m^{-0.5}$. Some
popular multi-mass dynamical models for globular clusters, the
so-called Michie-King models, have this scaling built in by assumption
\citep{gunn} and have been used in several studies (e.g. in recent
years: \citealt{paust2010,beccari,maccarone11,sollima12}).

We have shown here that this paradigm is {\it incorrect}, using direct
N-body simulations which are free of any assumption regarding energy
equipartion. The luminous stars in a globular that has evolved for a
long time converge toward $\sigma \propto m^{-\eta}$, with
$\eta_{\infty} \approx 0.08 \pm 0.02$ independent of position in the
cluster. In this state, the velocity dispersion is nearly independent
of mass. The inner regions can reach values up to $\eta_{\rm max}
\approx 0.2$ after several initial half-mass relaxation times. Either
way, the luminous stars in globular clusters with realistic IMFs
always remain far from complete equipartition.

The physical mechanism that explains why some systems cannot attain
energy equipartition, namely the \citet{spitzer69} instability for a
two-component system, has been know for a long time.  Also,
\citet{vishniac78} showed that for a continuous mass function of the
Salpeter form, energy equipartition is not attainable. However, the
implications of these results have gotten little attention in the
literature. This is probably because previous investigations focused
solely on the stability of systems with respect to energy
equipartition (e.g., \citealt{spitzer69,vishniac78,kondratev}) or the
implied mass segregation. The latter is a consequence of energy
equipartition but depends on other things as well
(e.g. \citealt{khalisi07,gill08,pas09}).

No previous theoretical study appears to have addressed the dynamical
evolution of realistic star cluster models with a specific focus on
mass-dependent kinematics, to determine exactly how close or far one
expects systems to be from energy equipartition. We therefore made
this the topic of the present paper. This analysis is particularly
timely, because with the advent of HST proper motion studies, it has
now become possible to actually measure the relation between $\sigma$
and $m$ in real clusters \citep{anderson}. Measurements of this
relation for main-sequence stars in some two-dozen clusters are
forthcoming \citep{bellini}. This opens up a whole new discovery
space. Detailed data-model comparisons of mass-dependent kinematics
have the potential to shed new light on globular cluster dynamics and
on their evolution. 

We analyzed the stellar dynamics in the direct N-body simulations
previously carried out by \citet{trenti10}. We quantified the relation
between $\sigma$ and $m$, as function of time and position in the
cluster, for realistic IMFs and initial conditions. We find that this
relation is well-fit by a power-law of the form $\sigma \propto
m^{-\eta}$, both for single main-sequence stars and for compact
remnants. Compact remnants tend to have higher $\eta$ than
main-sequence stars (but still $\eta < 0.5$), due to their steeper
(evolved) mass function. In the present paper we have focused mostly
on the main-sequence stars, which are actually observable. Our main
conclusions from this analysis are as follows:

\medskip

\noindent (1) The value of $\eta$ generally increases linearly in the
first few initial half-mass relaxation times ($t_{rh}(0)$). The
increase is faster for Lagrangian bins closer to the center, where
relaxation times are shorter. The central bin (containing the inner
10\% of the stars as seen in projection) reaches a maximum $\eta_{\rm
  max} \approx 0.15 \pm 0.03$. The increase to this value is mostly
completed by $t \approx 3 t_{rh}(0)$. At large times, all radial bins
convergence on an asymptotic value $\eta_{\infty} \approx 0.08 \pm
0.02$.

\medskip

\noindent (2) No simulated system ever reaches a state close to
complete equipartition with $\eta=0.5$. Even our most favorable
conditions for equipartition to develop in the core, that is a steep
IMF and low retention fraction of remnants, still yield only
$\eta_{\max} \lesssim 0.19$. Also, restricting the analysis to
particles at the center of the system in a three dimensional, rather
than projected, sense does not change our conclusions (although this
does yield a slight increase $\Delta \eta \approx 0.05$).

\medskip

\noindent (3) Partial energy equipartition develops in an overall
strikingly similar way across all our numerical experiments, despite
the variety of initial conditions employed (compare, e.g.,
Figures~\ref{fig:eta_evol}, \ref{fig:eta_time_64k_vanilla}, and
\ref{fig:eta_evol_IMBH}). The maximum and asymptotic $\eta$ values do
not differ much between runs (e.g., see Table~\ref{tab:sim} and
Figs.~\ref{fig:eta_half} and~\ref{fig:eta_inf}). Some trends are
present depending on IMF and the content of compact remnants and
binaries, and those trends can be understood based on simple physical
arguments (see Section~\ref{subsec:interp}).

\medskip

\noindent (4) The simulation with a central IMBH has the least amount
of equipartition (as measured by $\eta_{\rm max}$) among the sample of
initial conditions considered. This result is consistent with the
suppression of mass segregation that we have observed in simulations
with a central IMBH \citep{tre07b,gill08,pas09}. Further
investigations with a larger sample of runs (especially with an IMBH)
are required to fully characterize the generality of this
result. Either way, this does suggest a new method for constraining
the possible presence of an IMBH in a globular cluster, namely through
the slope $\eta$ of the $\sigma$--$m$ relation.\footnote{Searching for
  an equipartition signature from an IMBH in the cluster core has the
  potential to provide a diagnostic at earlier times in the life of a
  system than does mass segregation, since mass segregation is a
  consequence of equipartition. In fact, $t\gtrsim 5~t_{rh}(0)$ is
  required to discriminate systems with or without an IMBH from mass
  segregation \citep{gill08}, while potentially a signature in $\eta$
  can be seen already at $t\gtrsim 2~t_{rh}(0)$.} This is completely
independent from methods based on the radial velocity dispersion
profile $\sigma(r)$ (e.g. \citealt{vdm10}), but can be pursued with
similar (proper motion) datasets.
 
\medskip

\noindent (5) Any existing results derived from dynamical modeling
that assumed complete energy equipartition by construction may be
affected by unknown biases that will need to be carefully
evaluated. For example, past studies have often relied on multi-mass
Michie-King models (e.g., \citealt{demarchi2000,beccari}), and the
underlying equipartition assumption of such models does not appear to
be correct. It would be worthwhile for future studies to quantify any
biases that might be introduced, or to modify the underlying model
assumption to $\sigma \propto m^{-\eta}$ with $\eta \not= 0.5$.

\medskip

\noindent (6) Comparison of our simulations to a measurement of $\eta$
from HST proper motions in the core of Omega Cen yields good
agreement. This supports the view that globular clusters are not
generally in energy equipartition.

\section*{Acknowledgments}
The authors are grateful to Jay Anderson and Andrea Bellini for
collaboration on related projects, and for providing observational
drivers for this theoretical work. Andrea Bellini kindly provided his
latest proper motion results on Omega Cen, which were compared to our
simulations in Section~\ref{sec:omegacen}. The authors thank Adriano
Agnello for useful suggestions and discussions. Support of Hubble
Space Telescope Theory proposals HST-AR-11284 and HST-AR-12156 was
provided by NASA through grants from STScI, which is operated by AURA,
Inc., under NASA contract NAS 5-26555. MT acknowledges support from
the Kavli Institute for Theoretical Physics, through the National
Science Foundation grant PHY05-51164. This research was supported in
part by the National Science Foundation through TeraGrid resources
provided by the National Center for Supercomputing Applications
(grants TG-AST090045 and TG-AST090094). This project is part of the
HSTPROMO collaboration,\footnote{For details see HSTPROMO home page at
  \url{http://www.stsci.edu/~marel/hstpromo.html}} a set of HST
projects aimed at improving our dynamical understanding of stars,
clusters, and galaxies in the nearby Universe through measurement and
interpretation of proper motions.


\label{lastpage}

\end{document}